# Predicting Situation Awareness from Physiological Signals

Kieran J. Smith, Tristan C. Endsley, Torin K. Clark

*Abstract*—Situation awareness (SA)—comprising the ability to 1) perceive critical elements in the environment, 2) comprehend their meanings, and 3) project their future states—is critical for human operator performance. Due to the disruptive nature of gold-standard SA measures, researchers have sought physiological indicators to provide real-time information about SA. We extend prior work by using a multimodal suite of neurophysiological, psychophysiological, and behavioral signals, predicting all three levels of SA along a continuum, and predicting a comprehensive measure of SA in a complex multi-tasking simulation. We present a lab study in which 31 participants controlled an aircraft simulator task battery while wearing physiological sensors and responding to SA 'freeze-probe' assessments. We demonstrate the validity of task and assessment for measuring SA. Multimodal physiological models predict SA with greater predictive performance ($Q^2$ for levels 1-3 and total, respectively: 0.14, 0.00, 0.26, and 0.36) than models built with shuffled labels, demonstrating that multimodal physiological signals provide useful information in predicting all SA levels. Level 3 SA (projection) was best predicted, and level 2 SA (comprehension) was the most challenging to predict. Ablation analysis and single-sensor models found EEG and eye-tracking signals to be particularly useful to predictions of level 3 and total SA. A reduced sensor fusion model showed that predictive performance can be maintained with a subset of sensors. This first rigorous cross-validation assessment of predictive performance demonstrates the utility of multimodal physiological signals for inferring complex, holistic, objective measures of SA at all levels, non-disruptively, and along a continuum.

## I. INTRODUCTION

WITH rapid advancements in automation, modern day operators—like drivers, pilots, and firefighters—face an operational paradigm shift, away from roles that perform a single task and towards ones that must monitor one or more automated tasks [1-2]. For instance, autonomous vehicles demand that drivers continuously monitor their surroundings for hazards and prepare to take control when necessary [3]. This shift from human-in-the-loop to human-on-the-loop control places new stressors on operators' situation awareness (SA) [4-7], leaving potential for serious accidents.

To overcome these new challenges, researchers have suggested that, just as humans must monitor automated systems and adapt their behavior accordingly, automated systems should monitor a human's SA and adapt in turn [8], [9]. SA is defined as 1) an operator's perception of mission-critical elements in the environment, 2) their comprehension of the meaning of those elements, and 3) their ability to project future states [10]. These three levels of SA capture unique cognitive processes and are prescriptive of different interface adaptations. However, measuring SA in real-time in real-world situations is challenging. Current measures of SA that capture all three levels were primarily designed for use in simulators. Therefore, many require experimenters to pause an operator's task and ask 'freeze-probe' questions about current and future states [10], [11]. This process is infeasible in many real-world operational scenarios. For this reason, researchers have sought additional methods to capture SA during task performance without interrupting operations by demanding active operator input.

### A. Related Work

One such option lies in physiological signals. Passive capture of psychophysiological, neurophysiological, and behavioral signals, can provide valuable insight into an operator's SA to an adaptive system without requiring any active input from an operator [12-19]. While predicting a state of knowledge like SA from physiological data streams may seem farfetched, a large body of foundational work has demonstrated strong descriptive associations between a broad array of physiological signals and various measures of SA [20], [21-25]. Furthermore, prior work has found preliminary and promising success in predicting SA with data streams including, but not limited to, eye tracking (EYE), electrocardiogram (ECG), respiration (RSP), electrodermal activity (EDA), electroencephalography (EEG), and functional near-infrared spectroscopy (fNIRS) [16-19, 26-27].

However, for physiological models to be implemented in future human-computer interfaces, additional research is needed to understand what is possible and how best to predict SA from physiology. While promising, prior approaches have used either single- or dual-sensor physiological montages [16-19, 26-28], despite a broad array of sensors showing predictive

This research is funded by a Draper Scholarship and a National Science Foundation Graduate Research Fellowship. Any opinion, findings, and conclusions or recommendations expressed in this material are those of the authors and do not necessarily reflect the views of the National Science Foundation or The Charles Stark Draper Laboratory, Inc. (*Corresponding author: Kieran J. Smith*).

Kieran J. Smith is with the Aerospace Engineering Department, University of Colorado, Boulder, CO 80301 USA on a Draper Scholarship from The Charles Stark Draper Laboratory, Inc., Cambridge, MA 02139 (e-mail: kieran.smith@colorado.edu).

Tristan C. Endsley is with The Charles Stark Draper Laboratory, Inc., Cambridge, MA 02139 (e-mail: tendsley@draper.com).

Torin K. Clark is with the Aerospace Engineering Department, University of Colorado, Boulder, CO 80301 (e-mail: torin.clark@colorado.edu).

utility [28], leaving gaps in our knowledge as to how useful each sensor is and as to how accurate predictions could be with a full sensor suite.

Prior work has focused on binary classifiers of 'high' or 'low' SA [17, 26-27]. While these results are valuable, predictions of continuous measures of SA would allow for more fine-tuned interventions or adaptations from autonomous systems. For instance, work has suggested that AI teammates could adjust the number of tasks that are automated in order to mitigate SA decrements associated with autonomy-induced complacency [29]. Such a system could use predictions of continuous measures of operator SA to adjust the number of automated tasks along a continuum, gradually incorporating autonomy when SA increases and gradually removing autonomy when SA decreases. However, if an operator's SA falls critically low, such a system may want to reverse its operating mode by instead taking over tasks where an operator may be primed to make mistakes. Here, estimates of SA along a continuum—or, at a minimum, ternary predictions differentiating between low, medium, and high—could be critical to system function.

Lastly, there are certain limitations to the measures of SA in prior work that could restrict their utility in a human-autonomy teaming context. For instance, some prior work has explored unidimensional [19] or 2-dimensional [16, 18] measures of SA that exclude at least one level of SA. By excluding levels of SA in their measurements, these models exclude information that could be valuable to an adaptive system [30]. For instance, an adaptive display might adjust information content to help maintain operator SA [29], adding alerts to improve low perception scores, synthesizing information to improve low comprehension scores, or adding predictive displays to improve low projection scores.

Other work uses operator awareness of a single pre-defined hazard as a proxy for SA [19-20]. This is a valuable strategy to reduce the scope of SA and thus reduce measurement noise in proof-of-concept predictive models. However, using a pre-defined subset of information to model an operator's SA could fail to account for SA lapses in other aspects of the scenario. For example, operator awareness of an autonomous vehicle's navigation strategy may not be necessary in a controlled experiment but could be critical in a real-world scenario.

Prior work that uses holistic measures of SA tends to do so in simple tasks that may not generalize to the complex operational scenarios that exist in the real-world [15, 26-27]. These studies show the promise of predictive models of SA. However, complex multi-tasking scenarios are known to augment automation-induced complacency [31]—one of the greatest challenges to SA in partially-automated systems [32]. Models built on simple tasks may not capture SA decrements induced by real modern systems.

*B. Objectives*

This study seeks to build on past work in four ways. First, we capture signals from a multimodal suite of six research-grade physiological sensors (detailed in Section II.B. Physiological Signals). Through this, we aim both to capture additional variance in SA and to evaluate comparative utility of different sensors. By doing so, we provide future investigations insight into which physiological sensors provide the most valuable information for prediction, enabling future work to reduce the burden of applying, wearing, and processing multiple sensors.

Second, we fit multiple linear regression models to predict each of the three levels of SA along a continuum from low to high SA. This retains the prescriptive utility of each level of SA [30] and enables future autonomous systems to provide nuanced, informed, and fine-tuned responses. Furthermore, we retain explainability of final models.

Third, we evaluate all three levels of SA in a complex, human-in-the-loop multi-tasking paradigm designed to simulate the challenges that many operators face. We present an objective SA assessment that captures a complex and holistic measure of SA by querying 91 unique questions covering all important task elements.

Lastly, we aim to follow prior work by rigorously evaluating model performance on unseen observations. Physiological responses can vary widely even within individuals, and models fit to one set of data may be overfit to training data or may not generalize in a predictive manner. With the high-dimensional datasets that physiological signals and human data collection tend to lead to, we rigorously evaluate our model-building strategy on null data to quantify its ability to produce false positives.

II. STUDY DESIGN

To generate predictive models of SA, we conducted a laboratory study in which participants engaged in a standard, broadly used aircraft simulation. The simulation required participants to maintain awareness of and manipulate elements related to four separate subtasks. We generated a rich dataset, including three-level, objective measures of SA and a six-sensor suite of multimodal physiological data. All procedures were approved by the University of Colorado Boulder Institutional Review Board.

*A. Participants*

Written informed consent was acquired from 31 participants (16 female) aged 20-38 (M=25.6, SD = 3.9), recruited through flyers and email lists at the University of Colorado Boulder. Participation was restricted to individuals with no history of seizures, no known parabens allergy, no color-vision deficiencies, no consumption of alcohol in the six hours preceding the study, and to those able to read the screen without glasses. Participants identified as a majority white and non-Hispanic, except for four participants who identified as Asian (2), mixed-race, and African-American respectively. Five participants had been diagnosed with Attention Deficit Hyperactive Disorder, of which three were taking medication. One participant was left-handed. Seven participants had prior experience flying aircraft, an additional ten had experience with flight simulators of varying fidelity, and a further additional seven had some experience with aerospace displays. Seven participants had no experience with either.

*B. Physiological Signals*

Each participant was equipped with the same suite of physiological signals. A BIOPAC MP160 was used to capture ECG, EDA, and respiration data. ECG was captured using 3 disposable electrodes, placed on the skin beneath right clavicle, left clavicle, and lower left abdomen, and capture of R-peaks was visually confirmed. EDA was recorded using two electrodes, placed on the instep of the left foot [33-34]. Skin conductance responses were visually confirmed by having participants take a deep breath and hold it [35]. Respiration was recorded using a strain gauge belt, worn around the sternum. The NIRx NIRSport2 was used to record fNIRS in an 8-source 7-detector montage across the frontal lobe. Neuroelectrics' Enobio captured EEG signals in a sparse 16-electrode montage covering frontal, central, parietal, and occipital locations. The specific montages for these sensors are included in Supplementary Materials (Appendix A). Lastly, Pupil Labs Pupil Core eye-tracking glasses recorded eye movements and pupillometry. Physiological signals were synchronized to one another and to task events using Lab Streaming Layer [36].

In our laboratory experiment, we employed this broad suite of physiological sensors which, in operational settings, may be too obtrusive. However, by collecting data from each sensor stream, we can determine those most critical in order to reduce sensor burden.

*C. Aircraft Simulator Task Battery*

The Multi-Attribute Task Battery II (MATB) is a set of generalized piloting tasks, designed by NASA to enable study of flight-deck relevant performance and human-autonomy teaming in non-pilot participants [37]. The MATB has found broad use in evaluating SA [38-41]. The MATB is divided into four separate subtasks, system monitoring, tracking, communications, and resource management. In system monitoring, participants attend to four dials and two lights, and click on each when it deviates from predefined 'nominal' conditions. In tracking, participants monitor an autopilot system and, when the autopilot fails, take manual control with a joystick to maintain a blue target within a dotted square. In communications, participants respond to audio files by changing the specified radio to the specified frequency, only when their own call-sign is referenced. Lastly, in resource management, participants maintain fuel levels in two tanks (A and B) by turning on and off pumps 1-8 to move fuel from four other tanks. Two additional 'informational' panels exist on the right side. The scheduling panel tells participants when they can expect self-directed audio files, and the pump status panel displays the current flow rate of each pump.

In addition to the standard MATB, we incorporate a series of modifications, designed to make the task more realistic and to expand the possibility for comprehension and projection questions (levels 2 and 3 SA). Towards this, we instructed participants that systems F5 and F6 represent the primary and secondary flight computers and that successive failures of both computers would lead the autopilot (tracking task) to disengage. Similarly, if multiple other aircraft were detected nearby (in the form of audio files directed to other aircraft), the autopilot would disengage. These modifications simulate limitations of modern automated systems. Systems F1-F4 represented the power levels for pumps 1-4 respectively, and deviations from nominal power levels would cause temporary pump failures. Lastly, participants were taught nominal flow rates for each pump but told that flow rates could change.

Participants in this study completed a single 24-minute simulation, divided into twelve trials by SA freeze probes. Trial lengths were randomized between 1.5 and 2.5 minutes so that participants could not predict the timing of each freeze probe. Participants were all instructed on the four subtasks using a standard slide deck, before completing a 10-minute practice trial, during which they were told to ask questions and given notes on elements of the task when confused. Brief quizzes presented shortly afterwards confirmed that each participant understood the training.

Unique event files were generated for each participant using a custom MATLAB script so that each participant experienced a different sequence of events. This also served to ensure that event-specific physiological responses, such as rapid refixations to a specific sequence of failures, would not appear in models. In the script, each trial was assigned one of two task load levels, one of two time-delays, and one of two motivation levels. These manipulations were designed to induce a range of SA. High and low task load trials contained 9 and 18 'primary' events—events with no explicit cause—respectively. Secondary events, or events caused by other events, were computed following the rules mentioned above. Time delays between primary and secondary events were either 2-5 seconds or 12-15 seconds, in order to differentially load working memory. Motivation levels will be discussed in the next section.

Participants were instructed that they would be monetarily penalized for every second that a task element spent in an off-nominal state, including systems F1-F6, tanks A and B fuel levels, radio frequencies, and the tracking target.

*D. Questionnaires*

The primary response variable from this study—the SA scores—came from an 18-question SA freeze-probe assessment. Following each trial, the task was paused, display elements were blanked, and yes-or-no questions were randomly selected from a bank of 91 questions. Questions were developed specifically for this work, using a goal-directed task analysis of the modified MATB [42] (available in Supplementary Materials, Appendix B) and targeted each of the 3 levels of SA (i.e., 6 questions per level). Binary response questions were employed to maintain a similar difficulty level across questions, since the primary goal of this work was to predict changes to an operator's overall SA, not to evaluate SA related to individual task elements.

For example, a level 1 SA (perception) question might ask an operator whether a given pump in the resource management subtask is switched on, requiring only up-to-date knowledge of the element's current state. A level 2 SA (comprehension) question might ask whether the fuel level of a given tank pump in the resource management subtask is increasing or decreasing. To answer correctly, this question requires that participants know the current state of each pump that interacts with the tank as well as how these pumps interact to affect the flow rate of the tank. Lastly, a level 3 SA (projection) question might ask a participant if they expect the tank to be in violation of its lower limit in the next 30 seconds, asking participants to use their comprehension of the current state to forecast future events.

Participants earned a monetary reward for each SA question they answered correctly. Prior to each trial, participants were told whether they would earn a normal reward on the upcoming post-trial SA assessment or whether that award would be doubled. The intention was to provide increased motivation for participants to maintain SA and to induce a broad range of SA scores.

In addition to SA assessments, we captured measures of working memory, engagement, mind-wandering, and workload. Results from these measures are outside the scope of this paper.

### III. Data Analysis

To predict SA, features were extracted from physiological responses during each trial and then associated with the SA scores from the freeze-probe questions for the same trial. A combined feature selection and model-building paradigm was used and then cross-validated to predict SA scores using physiological features. The following sections will detail the processing of physiological features, calculation of SA scores, and generation of models.

*A. Physiology*

Physiological signals were processed and features calculated using MATLAB R2023b. Each data stream underwent different cleaning based upon typical best practices in the literature. Data were then divided into epochs associated with 30-second pre-trial baseline periods, with full trial periods, with the final 20 seconds of each trial, and with a single 2-minute pre-experiment baseline period. Features extracted include standard and state-of-the-art, a full list of which is described below. For each feature, raw and normalized values were computed as well as values baselined by dividing out and subtracting out both pre-trial and pre-experiment baselines.

fNIRS data were cleaned and processed using NIRS Toolbox [43]. Raw data were resampled to 5 Hz, then converted to oxygenated and deoxygenated hemoglobin concentrations using the Beer-Lambert Law. Hemoglobin concentrations were band-pass filtered between 0.01 and 0.5 Hz before extracting epochs. Mean, variance, skew, kurtosis, slope, root-mean-square, area-under-the-curve, max amplitude, and time-to-max were then calculated over each epoch for both oxy- and deoxygenated hemoglobin.

EEG data were cleaned and processed using EEGLAB v2021.1 [44]. A band-pass filter (cutoffs at 0.5 and 50 Hz) removed baseline drift. Independent component analysis was run on the entire recording, and components associated with eye and muscle movements were removed from the signal. Power spectral density was calculated with a 4-second window and a 50% overlap, both overall and by region. Power bands extracted included delta (1-4 Hz), theta (4-8 Hz), alpha (8-13 Hz), beta (13-30 Hz), and gamma (30-50 Hz), and regions included frontal, medial, parietal, and occipital. An engagement index and a task load index were calculated from spectral bands [45]. Connectivity between frontal and occipital regions was calculated as the correlation between each frontal electrode and the average activity across three occipital electrodes.

Eye-tracking data were processed in Pupil Labs' Pupil Player software. Gaze points, fixations, blinks, and pupil diameters were calculated from raw data. MATLAB was then used to calculate blink rate, blink duration, average pupil diameter, and fixation duration. Recurrence quantification analysis was performed per epoch to generate measures of recurrence, laminarity, determinism, entropy, mean line length, and center of recurrence mass [46].

Processed data from both eye-tracking and EEG were combined to compute fixation-related potentials. EEG data in electrodes O1, O2, and POz were extracted from 200 milliseconds (ms) prior to the onset of each fixation to 300 ms following this onset. The first 200 ms were treated as a baseline, and the average of this period was subtracted out for each epoch. Then, all epochs from a given electrode were averaged. A P100 component was calculated as the maximum voltage between 0 and 200 ms, and an N170 component was calculated as the average voltage between the P100 peak and 300 ms. The coefficient of variation was computed across the baseline period (-200 to 0 ms) as well as the FRP period (0 to 200 ms), and the ratio of the two were included as a feature.

ECG data were band pass filtered from 1 to 100 Hz, with a Butterworth band stop filter at 60 Hz to remove US powerline noise. R-DECO was then used to identify and extract R-peaks from the data [47]. In MATLAB, we then calculated heart rate and various metrics for heart-rate variability including the successive difference between normal-normal intervals, the percentage of normal-normal intervals below 50, and the root mean square of successive differences.

Respiration data were band pass filtered between 0.05 and 3 Hz, and the MATLAB function findpeaks was used to identify the timing and amplitude of individual breaths. From these peaks, we computed respiration rate, median breath amplitude, and proxy measures for tidal volume and minute ventilation.

Lastly, EDA data were restricted to datapoints between -1 and 40 microsiemens (µS), and to slow moving changes (derivatives and inflections less than 0.5 µS/s or 0.5 µS/s2). A third order Savitzky-Golay filter was used to smooth and fill missing data. Ledalab [48-49] software decomposed EDA signals into tonic and phasic components using continuous decomposition analysis. Minimums, maximums, and means were calculated from both tonic and phasic components. Skin conductance responses were extracted and used to calculate the mean amplitude, sum of amplitudes, number, and area underneath these responses.

EEG and eye-tracking data were lost for one participant due to a mistake in data collection. Eye-tracking data were lost for four trials of another participant due to an issue with data storage. In these cases and in scenarios in which non-number values were encountered (i.e., divide-by-zero errors), feature values were imputed. Where possible, a linear interpolation was performed within that participant to fill missing values. When all of a participants trials were missing a given feature, a 12-nearest neighbors classifier was used to impute said feature.

*B. Situation Awareness*

SA scores were calculated as the sum of correct responses to SA questions at each level. This meant that scores at each level could range from 0-6, though the binary nature of questions made scores less than 3 rare. Raw SA scores ranged from 1-6 with a mean of 4.87.

Since the goal of this work was to predict differences in SA across time and across participants—not across system elements—we adjusted scores for the difficulty of individual questions. To do this, we first computed the percentage of correct responses across all participants for each of question in the bank of 91 questions. If a participant got a question correct, they were awarded one point and penalized the percentage (0-1) correct for that question. If a participant got a question incorrect, they were just penalized the percentage correct for that question. Adjusted scores were then standardized before model-building. For each trial, a total SA score was calculated as the sum of the adjusted SA scores at levels 1, 2, and 3.

Lastly, since freeze-probe SA scores face challenges with single-trial sensitivity, we computed a moving-average of 3 SA scores [50] within each participant. To avoid edge effects, we removed the first and last trials of each participant.

When calculated in this manner, average-of-3 adjusted SA scores (computed from the 6 questions for each of levels 1, 2, and 3, asked after each trial's freeze-probe) ranged from -3.59 to 1.41. Total SA scores ranged from -5.67 to 2.82. As expected, average of three adjusted SA scores were strongly correlated with raw SA scores (Spearman's $\rho > 0.55$, $p \ll 0.01$), but yield a more robust, objective measure of SA [50]. Across 31 participants with (nominally) 10 measures each, data was lost for 10 trials, resulting in 300 total SA observations for each of level 1, 2, 3, and total SA.

*C. Models*

The following paragraphs detail our model-building approach and model-evaluation approach. In the results section, we use this model-building approach with all available data and report selected features and their associated coefficients. We use the model-evaluation approach to generate predictions of SA scores and report measures of predictive performance.

To generate multiple linear regression models (direct models) capable of predicting SA scores from physiological features, we utilized a relaxed least absolute shrinkage and selection operator (LASSO) regression algorithm, developed for physiological-based cognitive state estimation [51]. LASSO is efficient with high-dimensional data and it selects features best suited for continuous prediction in a linear regression model. However, LASSO has been found to select noise variables and can converge slowly on very high-dimensional data [52]. This method we employed here overcomes these challenges through a two-step process. First, for each level of SA (1, 2, 3, and total), LASSO was run 50 times, and any predictor selected by any run of LASSO (with λ at either 1 standard error or minimum standard error) was recorded. Next, to refine the features selected, another 50 iterations of LASSO were run, beginning with only those features that had been previously selected.

Feature sets selected with mean-square error at 1-standard error were identified. Any feature set with more than 40 predictors was eliminated. The remaining features sets were used to fit ordinary least squares regression models. Each regression model was evaluated using internal exhaustive leave-one-trial-out and leave-one-participant-out cross-validation metrics. Models with internal $Q^2$ metrics within 0.2 of that models' $R^2$ were considered not overfit and were considered further. If no models passed these criteria, no model was presented (i.e., no model converged). If more than one model met the criteria for not overfitting, then that with the highest $R^2$ value was presented as the final model. In order to provide a single descriptive model that best describes our dataset, the terms and coefficients of the final model presented here were generated using all available data.

To provide an estimate of model performance, this process was evaluated using 5-fold cross-validation. 20% of the data was withheld from feature-selection, model-fitting, and model-selection steps, and predictions on this unseen data were made using the final selected model. Each participant was represented in each fold. Predictions were then constrained to lie within the range of training data to avoid extrapolation. This work greatly benefited from access to the Alpine high performance computing resource at the University of Colorado Boulder [53].

Since model performance was evaluated using 5-fold cross validation, model predictions were generated using 5 different models. Descriptions (i.e., terms and coefficients) of these 5-fold models are not included, in favor of the single descriptive model mentioned above. These predictions were computed using 80% of the overall data and thus provide a conservative (pessimistic) estimate of the final model performance.

To determine the likelihood of random data producing our results, SA scores were randomly permuted 50 different times (i.e., dissociated with their corresponding physiological response data and associated with a different trial's), and models were generated in the exact same fashion. These shuffled-labels models are used as a baseline measure of 'poor' predictive accuracy to compare to our predictive models.

Since our feature selection method excluded certain sensors from final models, we employed a sensor-fusion models to capture contributions from each sensor. These generated separate predictions from each sensor and used internal fit ($R^2$ values on training data) to weight each prediction. Weighted predictions were summed to calculate final predictions.

To determine the predictive utility of each sensor, we evaluate the predictive performance of single-sensor models used in the creation of sensor-fusion models. We compare performance across different sensors to identify reduced sensor suites that could produce comparable results.

Additionally, we performed an ablation study in which we removed a single sensor at a time and generated new predictions. From these, we use decreases in predictive performance to identify sensors that contribute most strongly to predictive performance (i.e., sensors that cannot be removed without significant detriments to model performances).

Lastly, we generate and present a final model using the subset of sensors found to be most useful.

## IV. RESULTS

The following subsections detail performance of direct models, sensor fusion models, and ablation study results. Mean absolute errors (MAE) have been divided by the standard deviation of the outcome variable to enable comparisons between different outcomes and to lend some interpretability.

### A. Direct models

All direct models had a 100% convergence rate (CR; i.e., all 5 folds of cross-validation produced a model that met the criteria of internal $Q^2$ metrics within 0.2 of that models' $R^2$ define as not being overfit). Predictive performance of direct models varied by SA level (Table 1). Models predicting level 2 SA performed worst with an average $Q^2$ of 0.00 and an average MAE of 0.78 across all five folds. Models predicting total SA performed highest on predictive measures with an average $Q^2$ of 0.36 and an average MAE of 0.61.

On average, shuffled-labels models performed substantially worse than direct models on all metrics. $Q^2$ scores at each level were consistently below zero. Table 2 details a full list of performance metrics at each level as well as standard deviations across the distribution of 50 shuffled-labels models.

Of these shuffled-labels models, none outperformed direct models at any level of SA in either $Q^2$ or MAE (Table 3). As 50 Monte-Carlo iterations were generated, this suggests that no more than 2% of shuffled models could potentially perform similarly well as our direct models. Shuffled-labels models did converge to a solution approximately 75% of the time, suggesting that our model-selection method may may manage to return overfit models when signal-to-noise ratios are low.

Features from fNIRS, EEG, eye tracking, ECG, and respiration were selected for inclusion in final, descriptive (i.e., fit to all data) models. A single ECG feature (heart rate variability) was selected for level 1 SA models, and a single respiration feature (respiration rate) was selected in both level 2 and total SA models. Eye tracking features were selected for levels 1, 2, and total SA, and EEG and fNIRS features appeared in all four models. Fig. 2 denotes how many features were selected from each sensor in each model, and Appendix C (Supplementary Materials) details the full list of model predictors and coefficients.

### B. Sensor Fusion Models

Sensor fusion models outperformed direct models at each level of SA in both $Q^2$ and MAE. $Q^2$ scores were 0.21, 0.29, 0.34, and 0.41 respectively for levels 1, 2, 3, and total SA. Standardized MAEs were 0.70, 0.68, 0.64, and 0.60. All sensor fusion models converged in all cross-validate folds. Table 4 details the full list of performance metrics for each level of SA. Sensor fusion models also outperformed their shuffled labels counterparts in both $Q^2$ and MAE (Table 5). Unsurprisingly, performance was comparable across shuffled labels models at each level, with average $Q^2$ values of approximately -0.08 and average MAEs of at least 0.81. Shuffled models converged a fairly high percentage of the time (up to 100%). Due to the way predictions were combined by weighted mean (with nonconverging models weighted 0), if any one sensor generated a viable model, then that sensor could be weighted to provide predictions. Direct models with shuffled labels only failed to converge 25% of the time, meaning that we might expect all six sensors to fail $(0.25^6)$ 0.02% of the time. However, we saw nonconvergence on 4/200 models or 2% of the time, two orders of magnitude more than would be expected based on pure

### TABLE 1
PERFORMANCE METRICS FOR DIRECT MODELS AT EACH LEVEL

| SA Level | $Q^2$ | MAE$_{stdzd}$ | CR |
|---|---|---|---|
| 1 | 0.14 | 0.73 | 1 |
| 2 | 0.00 | 0.78 | 1 |
| 3 | 0.26 | 0.66 | 1 |
| Total | 0.36 | 0.61 | 1 |

### TABLE 2
PERFORMANCE METRICS FOR DIRECT MODELS FIT USING SHUFFLED LABELS WITH STANDARD DEVIATIONS ACROSS 50 MONTE CARLO SHUFFLES SHOWN IN PARENTHESES

| SA Level | $Q^2$ | MAE$_{stdzd}$ | CR |
|---|---|---|---|
| 1 | -0.25 (0.13) | 0.86 (0.05) | 0.75 (0.22) |
| 2 | -0.27 (0.13) | 0.89 (0.04) | 0.76 (0.22) |
| 3 | -0.24 (0.10) | 0.86 (0.04) | 0.77 (0.21) |
| Total | -0.27 (0.15) | 0.88 (0.06) | 0.70 (0.26) |

### TABLE 3
PORTION OF SHUFFLED-LABELS MODELS THAT PERFORMED BETTER THAN OR EQUAL TO DIRECT MODELS AT EACH LEVEL AND IN EACH METRIC

| SA Level | $Q^2$ | MAE$_{stdzd}$ | CR |
|---|---|---|---|
| 1 | 0 | 0 | 0.24 |
| 2 | 0 | 0 | 0.34 |
| 3 | 0 | 0 | 0.28 |
| Total | 0 | 0 | 0.28 |

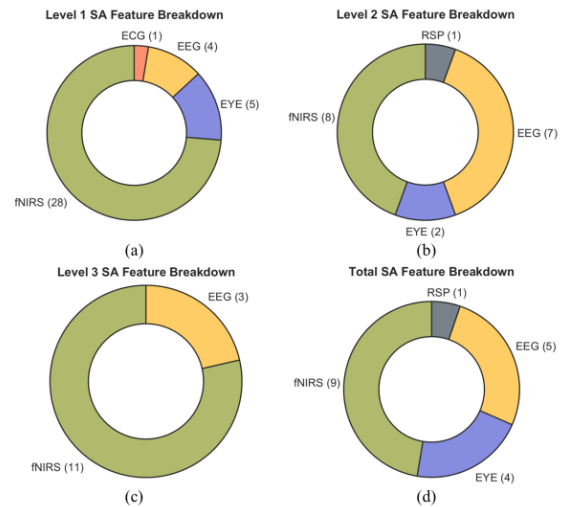

**Fig. 2.** Breakdown of features used in the descriptive model (using all data) for (a) level 1 SA (perception), (b) level 2 SA (comprehension), (c) level 3 SA (projection), and (d) total SA. Donut plots denote how many features came from each sensor.

### TABLE 4
PERFORMANCE METRICS FOR SENSOR FUSION MODELS AT EACH LEVEL

| SA Level | $Q^2$ | MAE$_{stdzd}$ | CR |
|---|---|---|---|
| 1 | 0.21 | 0.70 | 1 |
| 2 | 0.29 | 0.68 | 1 |
| 3 | 0.34 | 0.64 | 1 |
| Total | 0.41 | 0.60 | 1 |

TABLE 5
PERFORMANCE METRICS FOR SENSOR-FUSION MODELS FIT USING SHUFFLED
LABELS WITH STANDARD DEVIATIONS SHOWN IN PARENTHESES

| SA Level | $Q^2$ | $MAE_{stdzd}$ | CR |
|---|---|---|---|
| 1 | -0.09 (0.02) | 0.81 (0.05) | 0.96 (0.20) |
| 2 | -0.09 (0.02) | 0.84 (0.05) | 1.00 (0.00) |
| 3 | -0.08 (0.01) | 0.81 (0.05) | 1.00 (0.00) |
| Total | -0.08 (0.004) | 0.82 (0.04) | 0.99 (0.04) |

chance, suggesting that nonconvergence may occur more often with smaller (i.e., single sensor) feature sets.

Overall, we saw sensor fusion models outperform 100% of their shuffled-labels counterparts in both $Q^2$ and MAE (Table 6). Again, as 50 Monte-Carlo iterations were generated, this suggests that less than 2% of shuffled models would be expected to perform similarly to our predictive sensor-fusion models presented here.

Weightings used to generate final predictions were averaged across all five cross-validation folds to investigate which sensors were influencing predictions (Fig. 3). Neurophysiological sensors (EEG and fNIRS) and eye-tracking tended to be weighted higher than psychophysiological sensors (EDA, ECG, and respiration). For level 2 SA, however, EDA was weighted at 19% on average, on par with weightings for eye-tracking and EEG.

*C. Single Sensor Predictive Models*

Two-way ANOVAs were performed to analyze effects of sensor (EEG, fNIRS, eye-tracking, ECG, EDA, and respiration) and SA level (perception, comprehension, projection, and total SA scores) on both model performance metrics ($Q^2$ and MAE).

Both ANOVAs demonstrated similar results (Appendix D, Supplementary Materials). The two-way ANOVA for $Q^2$ revealed a significant interaction between effects of sensor and SA level (F(15, 119) = 2.33, p = 0.007) as well as significant main effects of sensor (F(5, 119) = 14.52, p = 0.00) and SA level (F(3, 119) = 4.59, p = 0.0048). The two-way ANOVA for MAE revealed a marginally significant interaction between effects of sensor and SA level (F(15, 119) = 1.58, p = 0.093) as well as significant main effects of sensor (F(5, 119) = 10.01, p = 0.00) and SA level (F(3, 119) = 3.15, p = 0.0286). These results suggest that certain sensors predict SA better than others but that this improved performance differs across levels of SA.

Pairwise comparisons were performed between different sensors using Tukey's honest significant difference. For $Q^2$, eye-tracking models were found to have significantly higher predictive performance than fNIRS (p = 0.0034), ECG (p = 0.0021), respiration (p = 0.0001), and EDA (p = 0.0001) models. EEG models were also found to have significantly higher predictive performance than fNIRS (p < 0.0001), ECG (p < 0.0001), respiration (p < 0.0001), and EDA (p < 0.0001) models. In terms of MAE, eye-tracking models had significantly greater predictive accuracy than ECG (p = 0.0053), respiration (p = 0.0007), and EDA (p = 0.0009) models as well as a marginally significant improvement over fNIRS models (p = 0.053). EEG models had significantly greater predictive accuracy than fNIRS (p = 0.0034), ECG (p = 0.0002), respiration (p < 0.0001), and EDA (p < 0.0001) models. No other significant differences were identified.

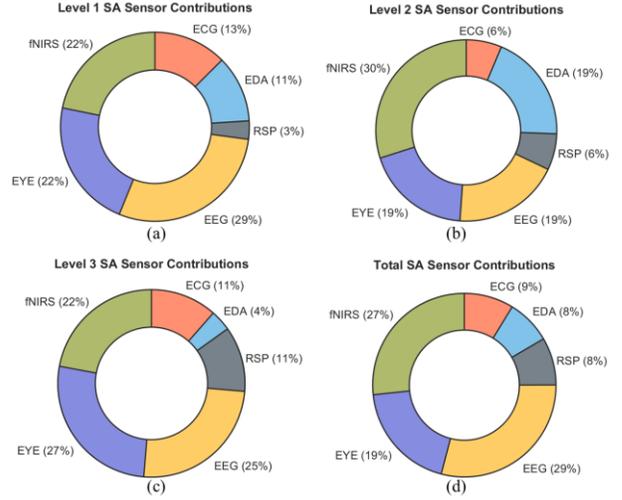

**Fig. 3.** Sensor fusion results for (a) level 1 SA (perception), (b) level 2 SA (comprehension), (c) level 3 SA (projection), and (d) total SA. Pie charts summarize average weighting of each sensor in the final model across each of five folds.

Predictive models using only eye-tracking signals had $Q^2$ scores of 0.13, 0.13, 0.23, and 0.23 respectively for levels 1, 2, 3, and total SA. Standardized MAEs were 0.73, 0.75, 0.66, and 0.69. Predictive models using only EEG signals had $Q^2$ scores of 0.10, 0.15, 0.35, and 0.37 respectively for levels 1, 2, 3, and total SA. Standardized MAEs were 0.74, 0.73, 0.67, and 0.61. These results are comparable to (and in some cases better than) direct models using all 6 sensors. However, sensor fusion models using all 6 sensors still outperform these single sensor models in terms of predictive accuracy metrics across all folds for each of level 1, 2, 3, and total SA.

*D. Ablation Study*

For the ablation study, single sensors were removed from the set of predictors one at a time and direct models were fit between predictors and each level of SA. To assess if predictive performance was degraded when removing each sensor, a one-way ANOVA was performed within each level of SA, but was only significant for total SA across $Q^2$ scores (F(6, 27) = 2.58, p = 0.042). Two-sample t-tests were performed between each ablation model and the total SA direct model (i.e., all sensors) presented in Results section A. For 6 total pairwise comparisons, p-values are Bonferroni adjusted. Total SA models performed significantly worse when EEG data was removed (t(8) = -5.10, p = 0.0056). Fig. 5 (Appendix E, Supplementary Materials) depicts $Q^2$ scores for each ablation model, with the direct model added for comparison. In summary, single sensor streams could be removed without dramatically degrading model predictive performance, except for EEG data for total SA models.

*E. Reduced Sensor Model*

Based on results from the single sensor predictive models and from the ablation study, it was determined that EEG and eye-tracking are significantly more predictive of SA and that model performance may deteriorate significantly without EEG signals. Therefore, a 2-sensor sensor fusion model was built using only eye-tracking and EEG signals. One participant was removed from this model as they were missing both eye-tracking and EEG data, leaving 30 participants.

These reduced sensor fusion models predicted levels 1, 2, 3, and total SA scores with predictive performances ($Q^2$) of 0.20, 0.16, 0.32, and 0.48 (MAEs = 0.71, 0.74, 0.64, and 0.63). Contributions of EEG signals to each model were around half, with some variation for each model (56%, 50%, 40%, and 40% for levels 1-3 and total, respectively).

## V. Conclusion

This study represents the first work, to our knowledge, to capture signals from a multimodal suite of six research-grade physiological sensors and to predict each level of SA along a continuum using multiple linear regression. We rigorously evaluated model performance on unseen observations and generated null (i.e., shuffled label) models for thorough comparison. Through this, we found find that our models pick up on genuine signals in our data and that physiological signals can provide useful information in predicting comprehensive objective measures of SA in complex tasks. We present results of a sensor ablation study and provide guidance to reduce sensor burden in future work.

### A. Direct models

Our finding that direct models outperformed all 50 shuffled models in both $Q^2$ and MAE suggests that physiological signals provide valuable information in the prediction of SA at all three levels (e.g., perception, comprehension, and projection). While this information content is evident at a scientific level, it may not yet be useful at an engineering level (i.e., operationally for real-time interventions). Predictive power ($Q^2$) never exceeded 0.36, and mean absolute errors remained above 0.6 when standardized. Future work is needed to investigate whether these models could benefit human-autonomy teams in real-time, and performance needs may depend upon the application.

### B. Sensor Fusion Models

Sensor fusion models appear to outperform direct models in predictive power and accuracy at each level of SA. This suggests that, while each sensor may not appear in a relaxed LASSO feature selection, these sensors still have predictive power to contribute to their final models. The contrast between direct models and sensor fusion models was most stark for level 2 SA, or comprehension, where $Q^2$ improved from 0.00 to 0.29, and MAE reduced from 0.78 to 0.68. It is possible that the complex underpinnings of comprehension particularly benefitted from an increased array of sensors. However, level 2 SA models performed the worst to begin with and thus may have simply had more room to improve.

### C. Sensor Burden Reduction

Results of the ablation study suggest that total SA models suffer a performance decrement without EEG signals. However, model predictive accuracy was robust to ablation of any other sensor stream, suggesting the other sensors were often able to compensate (i.e., provide similarly useful information for predicting SA). This offers an avenue for reducing sensor burden in operational environments or for tailoring sensor suites to particular environments. This work has limited sample size, and future work should evaluate what performance can be achieved with different subsets.

Single sensor predictive models suggest that EEG and eye-tracking signals may predict SA well at each level without additional sensors. Additionally, promising results of our 2-sensor sensor fusion model, using only eye-tracking and EEG signals, suggest that future work may not need to utilize the full physiological sensor suite described here to accurately predict objective SA scores. This enables future work to reduce burden on both experimenters and participants. That said, 2-sensor models did not outperform 6-sensor models at all levels of SA. Notably, level 2 SA predictions could have benefitted from variance captured by fNIRS (per single-sensor models) and EDA (per sensor-fusion weightings).

These results agree with prior work studying relationships between EEG signals and SA [18], [19], [26] as well as between eye-tracking and SA [16], [18], [26]. Prior work assessing SA and SA aptitude has found utility in fNIRS [17], [38].

That said, the wet-electrode EEG sensors used in this study often took the greatest amount of time to apply to participants and have high sensitivity to motion artifacts [54]. These limitations make it challenging to implement EEG signals in dynamic operational scenarios, where an operator cannot be seated still at a workstation. Future work should evaluate additional modeling strategies to understand relative importance of other sensors. Novel dry-electrode EEG systems that require less setup and have greater resistance to motion artifacts could also be evaluated in predictive SA models. Future work could aim to identify which of the EEG electrodes are most necessary out of our 16-electrode montage, which could also help reduce sensor burden and enable use in operational environments.

### D. Limitations

The work presented here used a goal-directed task analysis to generate SA assessment questions [55], however, the use of forced-choice binary questions to assess SA—as opposed to more traditional mix of binary, multiple-choice, and open-ended questions [55]—likely resulted in a noisier measure of SA. That said, binary questions have been used to evaluate SA in prior work [11, 56-57]. Furthermore, to emphasize sensitivity to the cognitive underpinnings of SA—over sensitivity to which elements of the task posed the greatest challenges to SA— this work attempted to account for question difficulty, and binary questions simplified this effort.

Each measure of SA consisted of only 6 binary questions at each level, creating a granulated measure of SA. We wished to assess each of levels 1, 2, and 3 SA on each trial and felt that asking more than 6 questions per trial would increase the amount of time between task-freeze and question presentation and, thus, introduce the potential for long term memory effects to confound SA. The moving average scores used here are designed to overcome limitations of the measures taken and have been shown to relate strongly with task performance [50].

Participants in the present work were not all trained experts in pilot-like tasks, and most had not been trained to gain and maintain SA. However, the task used in this work was designed to be simple enough to train participants to competency in a single visit, while maintaining ecological validity [37].

Lastly, despite our efforts to emphasize the cognitive underpinnings of SA, SA remains a contextual construct, and the features used here may be predictive of cognitive processes that are not used in the maintenance of SA in other tasks or domains. The MATB was designed to challenge many cognitive processes—like vigilance, planning, and multi-tasking—required in many operational settings, but care should be taken in extrapolating these findings to other domains.

*E. Future Work*

Future work to extend this research will build predictive models capable of capturing non-linear relationships between SA and physiology. While linear regression models enable predictions of SA along a continuum, other machine learning methods may capture more complex relationships between SA and underlying physiology. Furthermore, binary classifiers may present an 'easier' problem to begin with before advancing to predictions along a continuum.

Future models will incorporate relevant operator background information, such as training, prior experience, and sleep data into models. Broadly available in operator populations, and increasingly accessible in civilians through expanded use of wearable sensors and smart technology, this information could easily be used in real-world settings to complement physiological signals collected while performing a task.

The dataset reported here has been used to generate a 'cohort' model, or a single model for all participants. However, with broad variation between individuals both in physiology and in cognitive strategies for maintaining SA, it may be necessary to generate models fit or fine-tuned to a specific individual.

While feasible in a laboratory setting, donning six different physiological sensors may not be practical in many real-world settings. Future work could use this dataset to further determine relative utility of different sensors as a function of their unique burdens, and to use those results to reduce sensor burden.

Lastly, future work should aim to develop adaptive systems that might use these predicted SA scores to adapt in real-time to an operator, and evaluate whether such adaptations can improve SA, performance, or both. It is unclear at present how accurate these predictions must be to improve the overall performance of a human-autonomy team.

ACKNOWLEDGMENT

This research is funded by a Draper Scholarship and a National Science Foundation (NSF) Graduate Research Fellowship. Any opinion, findings, and conclusions or recommendations expressed in this material are those of the authors and do not necessarily reflect the views of either organization. The authors would like to thank Sofia Ibarra and Jasper Shen for their contributions to experimental setup and data collection. The authors would like to thank Dr. Troy Lau for providing insight and expertise that assisted in this research.

This work utilized the Alpine high performance computing resource at the University of Colorado Boulder. Alpine is jointly funded by the University of Colorado Boulder, the University of Colorado Anschutz, Colorado State University, and the NSF (award 2201538).

APPENDICES

*A. Physiological Signals*

Fig. 4 (Supplementary Material) depicts the scalp locations of all EEG electrodes and fNIRS optodes.

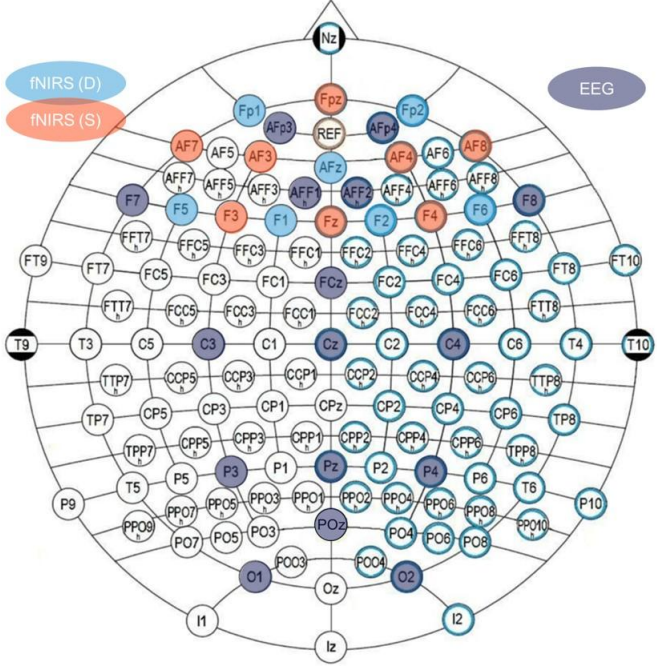

**Fig. 4.** Scalp locations of EEG electrodes and fNIRS optodes.

*B. Situation Awareness Assessment Questions*

Table 7 presents a complete list of SA questions from which those presented to participants were randomly selected.

TABLE 7
Complete list of SA questions

| Level 1 | Level 2 | Level 3 |
|---|---|---|
| Did the plane receive any instructions from audio files in the last 30 seconds? | Is the F[5/6] warning light nominal? | If no changes to the pumps occur, do you expect tank [A / B] to be in violation of its [upper / lower] limit in 30 seconds. |
| Is the F[5/6] warning light [red/green]? | Is the F[1-4] warning scale nominal? | If no changes to the pumps occur, do you expect tank [C / D] to be empty in 30 seconds. |
| Is the F[1-4] scale centered? | Is the autopilot currently operating on the secondary flight computer? | Do you expect the autopilot to go manual (fail) soon (within 15 seconds)? |
| Is the tracking task in manual mode? | Will a comm session [begin / end] within the next 30 seconds? | Do you expect to have to change a radio frequency soon (within 30 seconds)? |
| Were you instructed to set the [NAV1 / NAV2 / COM1 / COM2] radio to a different frequency during the last 15 seconds? | Does the [NAV1 / NAV2 / COM1 / COM2] radio need to be changed? | Do you expect pump [1 – 4] to fail within the next 15 seconds? |
| Is a communication session active? | Are there several planes nearby? | If no changes to the pumps occur, do you expect pump [1, 3, 5, 6] to turn off within the next 15 seconds? |
| Is your plane's call sign NASA 504? | Is the fuel level of tank [A / B / C / D] increasing? | |
| Is pump [1-8] currently [turned on / failed / turned off]? | Is the flow rate for pump [1-8] nominal (when switched on)? | |
| Is the fuel level of tank [A/B] within the light blue range? | Would turning [on / off] pump [1 – 4] help to minimize any current tank violations? | |
| Is the fuel level of tank [C / D] [more / less] than [¾ / ¼] full? | Do you currently need to turn on pump [5 / 6] to be able to use pump [1 / 2]? | |
| Is the flow rate for pump [1-8] [above / below] [800 / 500] (when switched on)? | | |

*C. Model Coefficients*

Tables 8-11 (Supplementary Material) show the features included in each model (Levels 1, 2, 3, and Total SA respectively). The first column denotes the feature, the second column denotes which sensor that feature was computed from, and the third column denotes the epoch during which this feature was computed. The corresponding coefficients show the linear regression fitted slope relating each feature to the prediction of adjusted SA. For example, a total SA score could be predicted by computing

$$SA_t = 40.02 - 0.29 * X_1 + 0.31 * X_2 - 0.01 * X_3 + ... + C_n X_n \quad (1)$$

Where $SA_t$ refers to the adjusted, average of three, total SA score, $X_1$ represents respiration rate (from the respiration sensor) computed over the entire trial and standardized across trials, $X_2$ refers to the average pupil diameter (eye tracking) over the final 20 seconds of the trial, and $X_3$ refers to the determinism of fixations over the entire trial period. $C_n$, and $X_n$ refer to the final coefficient and variable of the n features selected by LASSO, respectively.

TABLE 8
COMPLETE LIST OF FEATURES INCLUDED IN LEVEL 1 SA MODEL

| Sensors | Feature | Epoch | Coeff. |
|---|---|---|---|
| Intercept | | | 4.73 |
| ECG | RMSSD | Final 20s | 0 |
| Eye Tracking | average pupil diameter | Trial | 0.5 |
| Eye Tracking | average pupil diameter | Final 20s | -0.3 |
| Eye Tracking | determinism | Trial | -0.01 |
| Eye Tracking | mean line length | Trial | -0.28 |
| Eye Tracking | recurrence | Trial / First Baseline | 0.13 |
| EEG | average relative Frontal alpha band power | Final 20s / First Baseline | -0.04 |
| EEG | average relative Frontal beta band power | Trial / First Baseline | -0.04 |
| EEG | p100 o1 | Trial / First Baseline | 0 |
| EEG | n170 poz | Trial - First Baseline | -0.11 |
| fNIRS | S1D2 HbR skewness | Trial Standardized | -0.06 |
| fNIRS | S1D2 HbR slope | Trial - First Baseline | 0 |
| fNIRS | S2D1 HbR area under the curve | Trial - First Baseline | 0 |
| fNIRS | S2D1 HbR area under the curve | Final 20s - First Baseline | 0 |
| fNIRS | S2D1 HbR mean | Final 20s - First Baseline | 0.01 |
| fNIRS | S2D1 HbR skewness | Final 20s - Pre-Trial Baseline | 0.01 |
| fNIRS | S3D1 HbR skewness | Trial / Pre-Trial Baseline | 0 |
| fNIRS | S3D1 HbR area under the curve | Final 20s - Pre-Trial Baseline | 0 |
| fNIRS | S3D4 HbR mean | Final 20s Standardized | 0.08 |
| fNIRS | S4D2 HbR kurtosis | Final 20s | -0.06 |
| fNIRS | S4D2 HbR slope | Trial | 0 |
| fNIRS | S4D5 HbR slope | Trial | 0 |
| fNIRS | S4D5 HbR max amplitude | Trial | 0 |
| fNIRS | S5D3 HbR skewness | Final 20s | 0.03 |
| fNIRS | S5D3 HbR skewness | Final 20s - First Baseline | -0.08 |
| fNIRS | S6D4 HbR time to max | Trial / Pre-Trial Baseline | -7.75 |
| fNIRS | S6D6 HbR area under the curve | Final 20s | 0 |
| fNIRS | S6D7 HbR slope | Final 20s Standardized | 0.09 |
| fNIRS | S7D5 HbR kurtosis | Final 20s | 0.01 |
| fNIRS | S7D5 HbR kurtosis | Trial | -0.01 |
| fNIRS | S7D5 HbR slope | Trial | 0.01 |
| fNIRS | S7D5 HbR area under the curve | Trial Standardized | 0.11 |
| fNIRS | S7D5 HbR time to max | Trial - Pre-Trial Baseline | 0 |
| fNIRS | S7D5 HbR time to max | Trial / Pre-Trial Baseline | 3.54 |
| fNIRS | S8D6 HbR mean | Final 20s | -0.01 |
| fNIRS | S8D6 HbR kurtosis | Final 20s - Pre-Trial Baseline | -0.03 |
| fNIRS | S8D6 HbR time to max | Trial - Pre-Trial Baseline | 0 |
| fNIRS | S8D7 HbR RMS | Final 20s / First Baseline | 0.1 |

TABLE 9
COMPLETE LIST OF FEATURES INCLUDED IN LEVEL 2 SA MODEL

| Feature | Sensor | Epoch | Coeff. |
|---|---|---|---|
| Intercept | | | 8.82 |
| Resp. | RR | Trial Standardized | -0.07 |
| Eye Tracking | blink duration | Trial - Pre-Trial Baseline | 0.04 |
| Eye Tracking | recurrence | Trial - First Baseline | 0.01 |
| EEG | average relative occipital alpha band power | Trial | 3.52 |
| EEG | average relative medial alpha band power | Trial | 1.24 |
| EEG | p100 O1 | Trial Normalized | -0.17 |
| EEG | n170 O2 | Trial - First | -0.06 |

| Feature | Sensor | Epoch | Coeff. |
|---|---|---|---|
|  |  | Baseline |  |
| EEG | n170 POz | Trial - First Baseline | -0.08 |
| EEG | n170 POz | Final 20s / Pre-Trial Baseline | -0.02 |
| EEG | FRP coefficient of variation O2 | Trial - First Baseline | 1.66e4 |
| fNIRS | S4D2 HbR kurtosis | Final 20s | -0.08 |
| fNIRS | S4D5 HbR max amplitude | Trial - Pre-Trial Baseline | 0 |
| fNIRS | S5D3 HbR max amplitude | Final 20s | 0 |
| fNIRS | S5D3 HbR max amplitude | Final 20s / First Baseline | -0.11 |
| fNIRS | S5D3 HbR time to max | Trial / Pre-Trial Baseline | -9.03 |
| fNIRS | S5D6 HbR RMS | Trial - First Baseline | 0 |
| fNIRS | S6D6 HbR mean | Final 20s | 0.02 |
| fNIRS | S7D7 HbR time to max | Trial Standardized | 0.12 |

TABLE 10
COMPLETE LIST OF FEATURES INCLUDED IN LEVEL 3 SA MODEL

| Feature | Sensor | Epoch | Coeff |
|---|---|---|---|
| Intercept |  |  | 16.7 |
| EEG | engagement index | Trial | -1.95 |
| EEG | average relative Frontal alpha band power | Trial / First Baseline | -0.42 |
| EEG | p100 poz | Trial / First Baseline | 0.02 |
| fNIRS | S2D3 HbR time to max | Trial / Pre-Trial Baseline | -20.54 |
| fNIRS | S4D2 HbR mean | Trial / Pre-Trial Baseline | 0.01 |
| fNIRS | S4D2 HbR slope | Final 20s / Pre-Trial Baseline | -0.13 |
| fNIRS | S4D4 HbR area under the curve | Final 20s - First Baseline | 0 |
| fNIRS | S4D5 HbR RMS | Trial | 0 |
| fNIRS | S4D5 HbR RMS | Final 20s | 0 |

| Feature | Sensor | Epoch | Coeff |
|---|---|---|---|
| fNIRS | S4D5 HbR max amplitude | Trial - Pre-Trial Baseline | 0 |
| fNIRS | S5D3 HbR RMS | Trial / First Baseline | -0.09 |
| fNIRS | S5D3 HbR time to max | Trial / Pre-Trial Baseline | 5 |
| fNIRS | S5D6 HbR area under the curve | Final 20s / Pre-Trial Baseline | 0.01 |
| fNIRS | S5D6 HbR skewness | Final 20s / First Baseline | 0.01 |

TABLE 11
COMPLETE LIST OF FEATURES INCLUDED IN TOTAL SA MODEL

| Feature | Sensor | Epoch | Coeff |
|---|---|---|---|
| Intercept |  |  | 40.02 |
| Resp. | RR | Trial Standardized | -0.29 |
| Eye Tracking | average pupil diameter | Final 20s | 0.31 |
| Eye Tracking | determinism | Trial | -0.01 |
| Eye Tracking | recurrence | Trial - First Baseline | 0.01 |
| Eye Tracking | recurrence | Trial / First Baseline | 0.51 |
| EEG | average relative Frontal alpha band power | Trial / First Baseline | -0.27 |
| EEG | average relative Frontal alpha band power | Final 20s / First Baseline | -0.37 |
| EEG | average relative medial alpha band power | Trial | 11.95 |
| EEG | p100 poz | Trial / First Baseline | 0.04 |
| EEG | n170 poz | Trial - First Baseline | -0.09 |
| fNIRS | S4D2 HbR max amplitude | Final 20s | 0 |
| fNIRS | S4D5 HbR RMS | Trial | 0.01 |
| fNIRS | S4D5 HbR max amplitude | Trial - Pre-Trial Baseline | 0 |
| fNIRS | S4D5 HbR max amplitude | Trial | 0 |
| fNIRS | S5D3 HbR RMS | Trial / First Baseline | -0.38 |

| Feature | Sensor | Epoch | Coeff |
|---|---|---|---|
| fNIRS | S5D3 HbR time to max | Trial / Pre-Trial Baseline | -19.13 |
| fNIRS | S6D4 HbR time to max | Trial / Pre-Trial Baseline | -21.43 |
| fNIRS | S6D6 HbR RMS | Final 20s | -0.02 |
| fNIRS | S7D5 HbR slope | Trial | 0.01 |

*D. Effects of Sensor and SA Level on Model Performance Metrics*

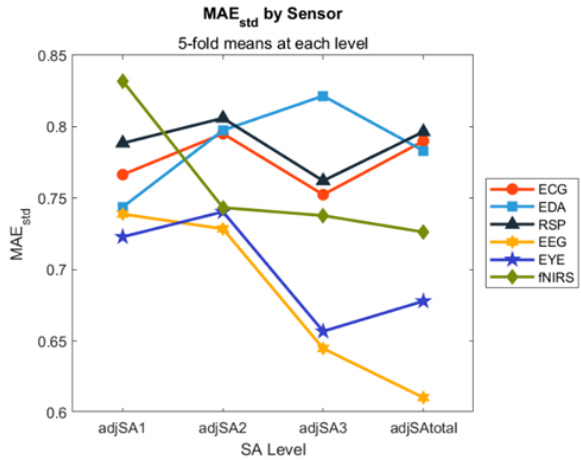

(a)

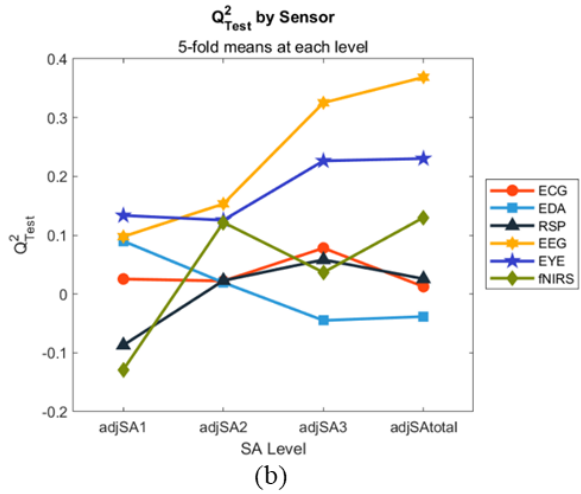

(b)

**Fig. 5.** Visualization of the significant interaction between effects of sensor and SA level on model performance metrics of (a) $Q^2$ (F(15, 119) = 2.33, p = 0.007) and (b) MAE (F(15, 119) = 1.58, p = 0.093)

*E. Visualization of Ablation Study Results*

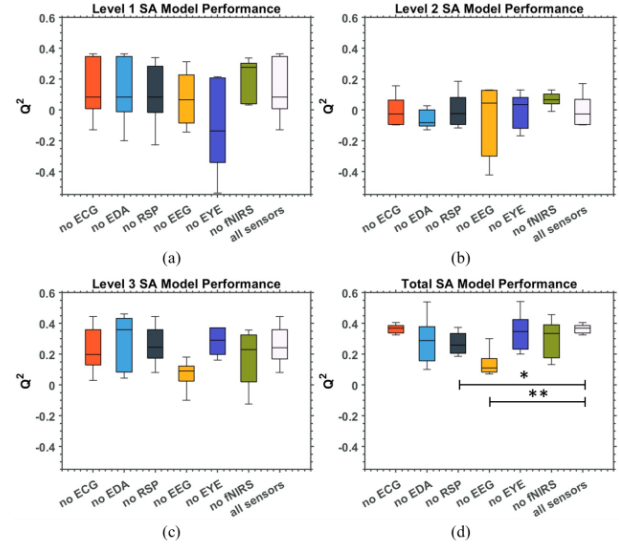

**Fig. 6.** Ablation study results for (a) level 1 SA (perception), (b) level 2 SA (comprehension), (c) level 3 SA (projection), and (d) total SA. Box plots summarize $Q^2$ scores across 5 cross-validation folds, and horizontal lines denote significance (*p = 0.05, **p = 0.0009).